\documentclass{amsart}
\usepackage{graphicx,cite,amsmath,amssymb}
\newtheorem{thm}{Theorem}[section]
\newtheorem{cor}[thm]{Corollary}

\theoremstyle{definition}

\theoremstyle{remark}

\def\beq{\begin{eqnarray}}
\def\eeq{\end{eqnarray}}
\def\bsp{\begin{split}}
\def\esp{\end{split}}

\def\d{\mathrm{d}}
\newcommand{\mf}[1]{{\mathfrak #1}}

\newcommand{\mb}[1]{{\mathbb #1}}
\newcommand{\mc}[1]{{\mathcal #1}}
\newcommand{\mbold}[1]{\mbox{\boldmath{\ensuremath{#1}}}}

\begin{document}

\title[\textbf{Negatively curved left-invariant metrics}]{\textbf{Negatively curved left-invariant metrics \\ on Lie groups}$^{\dagger}$}
\author{Sigbj{\o}rn Hervik}
\thanks{${}^{\dagger}$Talk held at the CMS/CAIMS Summer 2004 Meeting at Dalhousie University, Halifax, Nova Scotia, June 13-15.}
\address{Department of Mathematics \& Statistics, Dalhousie University, 
Halifax, Nova Scotia,
Canada B3H 3J5\newline (Current Address: University of Stavanger, N-4319 Stavanger, Norway) }%
\email{herviks@mathstat.dal.ca (\textrm{current:} sigbjorn.hervik@uis.no)}%
\date{\today}

\begin{abstract}
We discuss negatively curved homogeneous 
spaces admitting a simply transitive group of isometries, or equivalently, negatively curved left-invariant metrics on Lie groups. 
Negatively curved spaces have a remarkably rich and diverse 
structure and are interesting from both a mathematical and a
physical perspective. As well as giving general criteria for having
left-invariant metrics with negative Ricci curvature scalar, 
we also consider special cases, like Einstein spaces and 
Ricci nilsolitons. We point out the relevance these spaces 
play in some higher-dimensional theories of gravity. In particular, we show that the Ricci nilsolitons are Riemannian solutions to certain higher-curvature gravity theories. 
\end{abstract}
\maketitle 

\section{Introduction} 
Given a mathematical structure, some of the most interesting and basic mathematical objects are the 
symmetries of this structure. For a Riemannian space, $(M,{\bf g})$, we can consider the 
metric-preserving symmetries; i.e. the  \emph{isometries} defined by
\beq
\phi:~M\mapsto M, \quad \phi^*{\bf g}={\bf g}. \nonumber
\eeq
The isometries of a Riemannian space form a group, namely the \emph{isometry group}
\beq
\mathrm{Isom}(M)\equiv \left\{\phi:~M\mapsto M~|~\phi^*{\bf g}={\bf g}\right\}.
\eeq
The isometry group is a topological group, and, in fact, a \emph{Lie group} \cite{Kob}. This 
property of the isometry group makes the study of symmetries of Riemannian spaces interlinked with 
the theory of Lie groups. 

If two points, $p,q\in M$, are related via an isometry, i.e. $\exists\phi\in\mathrm{Isom}(M)$ such 
that $\phi(p)=q$, the metric at $p$ and $q$ will be mathematically indistinguishable. In a physical 
context this means that there is no measurement which can distinguish the two points. Hence, 
 the two points are equivalent and all physical properties (derived from the metric tensor) are 
 identical. 

A homogeneous Riemannian space $(M,{\bf g})$ is a space where the isometry group acts transitively;
 for any pair $p,q\in M$, there exist a $\phi\in\mathrm{Isom}(M)$ such that $\phi(p)=q$. Clearly,  
 homogeneous spaces have $\dim \mathrm{Isom}(M)\geq \dim(M)$. In the case where $\dim \mathrm{Isom}(M)=\dim(M)$
 and the action is transitive, we call the homogeneous space \emph{simply transitive}, 
 and if there there is a strict inequality we say the space is \emph{multiply transitive}. 
 It is common to write a connected homogeneous space as the space of cosets, $M=G/H$, where 
 $G$ the isometry group, and $H$ is a compact subgroup of $G$. For a more elaborate treatment 
 of Riemannian spaces and their isometry groups, see for example \cite{Kob}. 


\section{Homogeneous spaces and Lie groups}
Let us assume that we have a homogeneous Riemannian space $M=G/H$ which is connected and simply connected and
allows for a simply transitive
subgroup $K\subset G$ acting on $M$ so that $\dim(K)=\dim(M)=n$. This implies
that $M$ is a  manifold which can be identified as the group $K$. Hence, the
spaces we are considering are group manifolds, and, in fact, Lie
groups. Thus an equivalent point of view is to consider 
\emph{left-invariant metrics on Lie groups}. In the following we will use these two views 
interchangably. 

The connection between the Lie group structure of the manifold and the left-invariant Riemannian structure is most easily seen by using a left-invariant frame. The symmetries of the space can be expressed locally in terms of the Killing vectors. The simply transitive group of isometries can be taken to be the Lie group $K$ whose Lie algebra will be denoted $\mf{k}$. Then, there exist a ``Killing frame'' $\{{\mbold\xi}_i\}$ whose span is isomorphic to the Lie algebra of $K$; i.e. \footnote{Here and throughout the paper the Einstein summation conention has been used; i.e. summation over repeated indices.}
\beq
[{\mbold\xi}_i,{\mbold\xi}_j]=\widetilde{C}^k_{~ij}{\mbold\xi}_k,
\eeq
where the $\widetilde{C}^k_{~ij}$'s  are the structure constants of the Lie algebra $\mf{k}$. A left-invariant frame $\{ {\bf e}_i\}$ can now be found by requiring 
\beq
\pounds_{{\mbold\xi}_j}{\bf e}_i=[{\mbold\xi}_j,{\bf e}_i]=0.
\eeq
This left-invariant frame will also form a Lie algebra:
\beq
[{\bf e}_i,{\bf e}_j]={C}^k_{~ij}{\bf e}_k.
\eeq
The structure constants ${C}^k_{ij}$ are related to $\widetilde{C}^k_{~ij}$ in the following way: If we choose the frames $\{{\mbold\xi}_i\}$ and $\{ {\bf e}_i\}$ to coinside at \emph{one} point, then ${C}^k_{~ij}=-\widetilde{C}^k_{~ij}$ which implies that these Lie algebras are isomorphic. The dual one-forms, ${\mbold\omega}^i$, of the vectors  $ {\bf e}_i$  are also left-invariant (i.e. $\pounds_{{\mbold\xi}_j}{\mbold\omega}^i=0$) and obey
\beq
\d {\mbold\omega}^k=-\frac 12C^k_{~ij}{\mbold\omega}^i\wedge{\mbold\omega}^j.
\label{eq:domega}\eeq
These left-invariant one-forms are  manifestly invariant under the group action so a particularly convenient choice is to use $\{ {\mbold\omega}^i\}$ as our orthonormal Cartan co-frame. Henceforth we will assume that our frame is orthonormal so that our metric components are $g_{ij}=\delta_{ij}$. 

Given a Lie algebra $\mathcal{A}$ with structure constants $C^k_{~ij}$ we define the $GL(n)$-orbit,
\beq
\mathcal{O}_{GL(n)}(\mathcal{A})\equiv \left\{ \widetilde{C}^k_{~ij}~\big|~\widetilde{C}^k_{~ij}=({\sf M}^{-1})^k_{~l}C^l_{~mn}{\sf M}^m_{~i}{\sf M}^n_{~j}, ~~{\sf M}\in GL(n)\right\},
\eeq
 and the $O(n)$-orbit, 
\beq
\mathcal{O}_{O(n)}(\mathcal{A})\equiv \left\{ \widetilde{C}^k_{~ij}~\big|~\widetilde{C}^k_{~ij}=({\sf M}^{-1})^k_{~l}C^l_{~mn}{\sf M}^m_{~i}{\sf M}^n_{~j}, ~~{\sf M}\in O(n)\right\}.
\eeq
From the definition of a Lie algebra, two Lie algebras are isomorphic if and only if they lie in the same $GL(n)$-orbit $\mathcal{O}_{GL(n)}(\mathcal{A})$. Regarding the Riemannian structure, a frame rotation corresponds to an $O(n)$-transformation; i.e. two left-invariant frames whose structure constants lie in the same orbit $\mathcal{O}_{O(n)}(\mathcal{A})$ are isometric as Riemannian spaces. Since, $\mathcal{O}_{O(n)}(\mathcal{A})\subset \mathcal{O}_{GL(n)}(\mathcal{A})$, we can by varying the structure constants $C^k_{~ij}$ consider various  left-invariant Riemannian structures on the Lie group. In fact, all left-invariant Riemannian metrics on a given Lie group can in this way be obtained  by varying the structure constants, $C^k_{~ij}\in \mathcal{O}_{GL(n)}({\mathcal{A}})$, while keeping the metric fixed. 

Explicitly, in terms of $C^k_{~ij}$  the Ricci curvature can be written
\beq
R_{ij}=\mathcal{R}_{ij}-\frac 12\mathcal{K}_{ij}-\mathcal{Z}_{ij},
\label{eq:Rij}\eeq
where $\mathcal{K}_{ij}$ is the Killing form of $\mf{k}$, 
\beq
\mathcal{K}_{ij}=C^{a}_{\phantom{a}bi}C^b_{\phantom{b}aj},
\eeq
and $\mathcal{R}_{ij}$ and $\mathcal{Z}_{ij}$ are defined by
\beq
\mathcal{R}_{ij}&=&\frac 14 C_{iab}C_j^{\phantom{j}ab}-\frac 12C^{ab}_{\phantom{ab}i}C_{abj}, \nonumber \\
\mathcal{Z}_{ij}&=& C^a_{\phantom{a}ab}C^{\phantom{(ij)}b}_{(ij)}.
\eeq
In this way we see how the Lie algebra structure of the group is directly involved in the Riemannian structure. For various Lie groups the constants $C^k_{~ij}$ can only be of certain forms. For example, $\mathcal{Z}_{ij}=0$ for unimodular groups\footnote{A group is called \emph{unimodular} if it admits a bi-invariant measure. In terms of the structure constants, $C^i_{~ji}=0$.}  and $\mathcal{Z}_{ij}=\mathcal{K}_{ij}=0$ for nilpotent groups\footnote{A Lie algebra is nilpotent if the decending series ${\mathfrak g}_0\equiv {\mathfrak g}$ ${\mathfrak g}_i\equiv [{\mathfrak g}, {\mathfrak g}_{i-1}]$ terminates; i.e. there exists a $k$ such that ${\mathfrak g}_k=0$. A group is called nilpotent if and only if its Lie algebra is nilpotent.}. Hence, the group theoretical properties will to some extent be reflected in the Ricci curvature.

Let us first discuss some of the global aspects of Lie groups/Lie algebras. Any Lie group defines uniquely a Lie algebra; however, to any
given Lie algebra there may be many different Lie groups. For
example, the groups $SU(2)$ and $SO(3)$ are different but they have
isomorphic Lie algebras. Notwithstanding this, groups having
isomorphic Lie algebras are locally the same (\cite{Helgason} page
109): 
\begin{thm}
Two connected Lie groups are locally isomorphic if and only if their Lie
algebras are isomorphic.
\end{thm}
Hence, $SU(2)$ and $SO(3)$ can only differ at a global scale. Assuming
simply connectedness removes this ambiguity completely\footnote{Regarding global properties of Lie groups we refer the reader to the book \emph{Theory of Lie Groups} by Chevalley \cite{Chevalley}. Here, many classic results on the global structure of groups are given.} 
(\cite{Chevalley} page 113):
\begin{thm}
Let $G$ and $H$ be connected Lie groups with Lie algebras $\mf{g}$ and
$\mf{h}$, respectively. If $G$ is simply connected, then for every homomorphism
of $\mf{g}$ into $\mf{h}$ there exists a homomorphism of $G$ into $H$.
\end{thm}
This implies that if two connected and simply connected Lie groups
have isomorphic Lie algebras, then they must be isomorphic as Lie
groups. These results reduce the problem of classifying simply
connected groups $K$ to classifying their Lie algebras $\mf{k}$. 
 We will also assume that there is a Riemannian
structure on $M$ (which is not unique) given in terms of a $K$-invariant metric
${\bf g}$.  

\section{The Ricci curvature scalar}
A question one would like to answer is: Can we say anything about the Ricci curvature scalar, defined as $R=R^i_{~i}$, from the properties of the Lie group? In other words, to what extent does the curvature of left-invariant metric depend on the properties of the Lie group? 

Let us consider the sign of the Ricci curvature scalar of ${\bf g}$. First we ask: Which Lie groups \emph{admit} a left-invariant metric with negative Ricci scalar, $R< 0$? The answer is provided by the following theorem \cite{Milnor:76}:
\begin{thm}
If the Lie algebra of $K$ is non-commutative, then $K$ possesses a left-invariant metric of strictly negative scalar curvature. 
\end{thm}
\begin{proof}
The proof utilizes the following fact: For any non-Abelian Lie algebra $\mf{k}$, we have either of the two (mutually exclusive) 
possibilities
\begin{enumerate}
\item{} There exists a Lie algebra
contraction $\phi$ such that $\mf{k}\overset{\phi}{\longrightarrow} \mf{n}_3+\mb{R}^m$, where $\mf{n}_3$
is the 3-dimensional Heisenberg Lie algebra. 
\item{} Any left-invariant metric on $K$ has constant negative curvature. 
\end{enumerate}
The first implies that there exists a continuous curve $c(t)$ in the orbit $\mathcal{O}_{GL(n)}(\mf{k})$ which has 
$\lim_{t\rightarrow\infty}c(t)\in \mathcal{O}_{GL(n)}(\mf{n}_3+\mb{R}^m)$, and hence, we can come arbitrary 
close to the Lie algebra $\mf{n}_3+\mb{R}^m$. Since the Heisenberg group has strictly negative Ricci scalar curvature
there must exist, by continuity, a left-invariant metric on $\mf{k}$ with negative Ricci scalar curvature as well. 
The second case has trivially negative Ricci scalar curvature. 
\end{proof}
So if the group $K$ is not Abelian, $K$ admits a left-invariant metric such that $R<0$. This may be surprising to many, but this means that even compact groups -- e.g. $SU(2)\cong S^3$ --  admits negatively curved left-invariant metrics.

A natural follow-up would be to consider Lie groups for which \emph{all} left-invariant metrics have $R<0$. To find these the
following theorem by B\'erard Bergery \cite{BB:78}\footnote{This
  theorem is based on a conjecture by Milnor \cite{Milnor:76}.} comes
to our aid:
\begin{thm}[B\'erard Bergery] \label{thm:Rleq0}
Let $M=G/H$ be a homogeneous space on a connected Lie group $G$,
admitting a $G$-invariant metric. Then the following 2 statements
are equivalent:
\begin{enumerate}
\item{} The universal cover of $M$ is diffeomorphic to a Euclidean space.
\item{} Every $G$-invariant metric on $M$ is either flat or has strictly
  negative Ricci curvature scalar. 
\end{enumerate}
\end{thm}
This theorem is very powerful and gives us the necessary
criterion for non-positiveness of $R$. Firstly, the only
homogeneous manifolds whose universal covers are diffeomorphic to $\mb{R}^n$ and has $R=0$, are the flat ones. Secondly, it lifts the local statement
about the Ricci scalar to a statement of the global structure of
$M$. Hence, to check whether $M$ has a non-positive Ricci
curvature scalar we have to check the global structure of the group
$K$. 

Another helpful result is the Iwasawa decomposition \cite{Iwasawa}:
\begin{thm}[Iwasawa]\label{thm:Iwasawa}
If $K$ is a connected Lie group, then
\begin{enumerate}
\item{} Every compact subgroup is contained in a maximal compact subgroup H, which is necessary a connected Lie group.
\item{} This maximal compact subgroup is unique up to conjugation.
\item{} As a topological space, $K$ is homeomorphic with the product of $H$ and some Euclidean space $\mb{R}^m$: 
\[ K\cong H\times \mb{R}^m\quad\]
\end{enumerate}
\end{thm}
An immediate consequence is the following \cite{Milnor:76}: 
\begin{cor} \label{cor:Iwasawa}
The universal covering of $K$ is homeomorphic to Euclidean space if and only if every compact subgroup is commutative.
\end{cor}
This is exactly what we need to find out whether a Lie group has a universal cover diffeomorphic to a Euclidean space. 

\subsection{The classification of Real Lie algebras}

To check the global structure of $M$ it is necessary to remind ourself
of some classification results regarding real Lie algebras. We shall use
the symbol $\mf{s}\oplus\mf{h}$ for the semidirect sum, writing the
ideal $\mf{s}$ first, and the subalgebra $\mf{h}$ second (an ordinary
$+$ is written for the direct sum). Using
$[-,-]_{\mf{s}}$ and $[-,-]_{\mf{h}}$, we can endow the semidirect sum
with a Lie algebra structure as follows:
\beq
[e_i,e_J]=D(e_i)\ast e_J,\quad e_i\in \mf{h},\quad e_J\in\mf{s},
\eeq 
where $D(e_i)$ is a linear mapping, $D(e_i):\mf{s}\mapsto\mf{s}$. This
leads to the following relations
\beq
[\mf{s},\mf{s}]\subset\mf{s}, \quad [\mf{h},\mf{h}]\subset\mf{h},
\quad [\mf{s},\mf{h}]\subset\mf{s}. 
\eeq
Furthermore, the Jacobi identity implies that $D(e_i)$ is a derivation
of $\mf{s}$:
\beq
D(e_i)\ast [e_J,e_K]=[D(e_i)\ast e_J,e_K]+[e_J,D(e_i)\ast e_K].
\eeq
The set $\left\{D(e_i)\right\}$ is itself a Lie algebra, and the
homomorphism $e_i\mapsto D(e_i)$ must be a representation of the
algebra $\mf{h}$. The fundamental Levi-Malcev
theorem says that, for an arbitrary Lie algebra $\mf{g}$ with radical
$\mf{s}$, a semisimple subalgebra $\mf{h}$ exists such that 
\[ \mf{g}=\mf{s}\oplus\mf{h}.\]
The semisimple subalgebra $\mf{h}$ is called the \emph{Levi
factor}. This immediately implies that the Lie algebras fall in three
categories: 
\begin{enumerate}
\item{} The solvable algebras.
\item{} The semisimple algebras.
\item{} The semidirect sums of solvable and semisimple algebras.
\end{enumerate}
\begin{table}
\centering
\begin{tabular}{|c|c|c|c|} \hline 
dim & simple & semi-direct sum & solvable \\ \hline 
1 & & & 1 \\
2 & & & 1 \\
3 &2& & 7 \\
4 & & & 12 \\
5 & &1& 40 \\
6 &1&4& 164\\
7 & &7& ? \\
8 &3&22& ? \\
\hline     
\end{tabular}
\caption{The number of equivalence classes of indecomposable real Lie algebras up to dimension 8. These numbers should be used with care since some equivalence classes are actually continuous families of non-isomorphic Lie algebras.}\label{Tab:No}
\end{table}

\subsection*{The solvable algebras} Given a Lie algebra $\mf{g}$. The algebra $\mf{g}$ is called \emph{solvable} if the derived series $\mf{g}_i$, defined iteratively by  $\mf{g}_0=\mf{g}$ and $\mf{g}_{i+1}\equiv [\mf{g}_i, \mf{g}_i]$, terminates; i.e. there exists an $m$ such that $\mf{g}_m=0$. 

In spite of the fact that the solvable ones are the most numerous (see Table \ref{Tab:No}),
they are easy to deal with due to a result by \'E. Cartan \cite{Cartan}:
\emph{An $n$-dimensional group is diffeomorphic to $\mb{R}^{n}$ if it
is solvable, connected and simply connected}\footnote{See also \cite{Chevalley:41}. This paper also 
considers the more general
case when the solvable group is not necessary simply connected.}. In particular, this means that solvable algebras do not have any non-Abelian compact subgroups. More generally, if  $\mf{s}$ is solvable, then there does not exist a semisimple subalgebra
$\mf{h}$ such that $\mf{s}\supset\mf{h}\neq \{0\}$. 

A complete classification of the solvable Lie algebras exists only up to
dimension 6. In this regard, Turkowski found the remaining
ones of dimension 6 \cite{Turkowski:90}. (For the real and complex 7-dimensional nilpotent Lie algebras, see \cite{Magnin,Romdhani,Seeley}.) 

\subsection*{The semisimple algebras} 
 The study of
semisimple groups reduces to studying the simple groups. 
The real simple algebras come in two classes \cite{Helgason}: $(I)$, the simple Lie
algebras over $\mb{C}$, considered as real Lie algebras, and $(II)$,
the real forms of simple Lie algebras over $\mb{C}$. 

The semisimple Lie algebras over $\mb{C}$ are completely classified so
we can use these results. First, for the class $I$ real algebras note
that they \emph{all} have the subalgebra $\mf{s}\mf{l}(2,\mb{C})$ (by
inspecting, for example, the Dynkin diagrams). Due to the well-known
diffeomorphisms
\[ \widetilde{SL(2,\mb{C})}\cong SU(2)\times \mb{R}^3\cong S^3\times \mb{R}^3, \]
none of the class $I$ real Lie groups will be diffeomorphic to a
Euclidean space. The Lie algebras of class $II$ are slightly more
subtle. Using the Dynkin diagrams and the root systems, we  see
that all the Lie algebras over $\mb{C}$, except $A_1$ and $B_2$, have
a $A_2$-subalgebra. $A_2$ has the following real forms:
\beq
\widetilde{SL(3,\mb{R})}\cong SU(2)\times \mb{R}^5, \quad SU(3), \quad
\widetilde{SU(2,1)}\cong SU(2)\times \mb{R}^5.
\eeq
The algebra $B_2$ has the following real forms \cite{Helgason}:
\beq
SO(5), \quad \widetilde{SO(4,1)}\cong SU(2)\times SU(2)\times \mb{R}^4,
\quad \widetilde{SO(3,2)}=SU(2)\times \mb{R}^7. 
\eeq
Finally, the algebra $A_1$ has the following real forms:
\beq
\widetilde{SL(2,\mb{R})}\cong \mb{R}^3, \quad SU(2)\cong S^3. 
\eeq
Using Corollary \ref{cor:Iwasawa}, the above analysis implies that none of the simple Lie algebras, except for $\mf{s}\mf{l}(2,\mb{R})$, give rise to groups whose universal cover is diffeomorphic to a Euclidean space. 

\subsection*{Lie algebras which are semidirect sums}
For these algebras, there is one thing to note. Assume that
$\mf{l}=\mf{s}\oplus\mf{h}$, where $\mf{s}$ is solvable and $\mf{h}$
is semisimple and let $\widetilde{M}_{\mf{l}}$, $\widetilde{M}_{\mf{s}}$,
and $\widetilde{M}_{\mf{h}}$ be the universal cover of their
corresponding groups. Then we have the diffeomorphism
\beq
\widetilde{M}_{\mf{l}}\cong \widetilde{M}_{\mf{s}}\times \widetilde{M}_{\mf{h}}.
\eeq
Hence, the topology of these groups is the direct product of the
topology of their
respective components. Since  $\widetilde{M}_{\mf{s}}$ is
diffeomorphic to a Euclidean space, we can understand the groups in
this category by understanding the semisimple groups. 

All of these algebras are not known explicitly; however Turkowski has
found all the real Lie algebras having a  
non-trivial Levi decomposition up to dimension 9
\cite{Turkowski:88,Turkowski:92}. 

As an example,  consider the only non-trivial 5-dimensional Lie algebra of this type, which gives 
rise to the group $\mb{R}^2\ltimes SL(2,\mb{R})$. A left-invariant frame on this group can be taken to be
\beq
{\mbold\omega}^1=\d x+\cosh 2\theta\d y, \quad {\mbold\omega}^2=e^{-2x}\left(\d \theta+\sinh 2\theta\d y\right), \quad {\mbold\omega}^3=e^{2x}\left(\d \theta-\sinh 2\theta\d y\right)\nonumber \\
{\mbold\omega}^4=e^{-x}\left(e^{-y}\cosh\theta\d u-e^y\sinh\theta\d v\right), \quad {\mbold\omega}^5=e^{x}\left(-e^{-y}\sinh\theta\d u+e^y\cosh\theta\d v\right).\nonumber
\eeq
These left-invariant one-forms obey eq.(\ref{eq:domega}) where the $C^k_{~ij}$'s represent the Lie algebra $L_{5,1}$ in \cite{Turkowski:88}. 

\subsection{Negatively curved left-invariant metrics}
We can summarise our investigation in the following theorem. 
\begin{thm} Let $M=G/H$ be a connected $n$-dimensional homogeneous Riemannian
manifold admitting a 
  $G$-invariant metric. Assume further that the universal cover
  $\widetilde{M}$ of $M$ admits a  group $K$ acting simply
  transitive on $\widetilde{M}$. Then the following statements are equivalent:
\begin{enumerate}
\item{} Every $G$-invariant metric on $M$ has non-positive Ricci
curvature scalar. 
\item{} The Lie algebra of $K$ does not have a subalgebra isomorphic
  to the Lie algebra $\mf{s}\mf{u}(2)$.
\end{enumerate} 
\end{thm}
\begin{proof}
Note that all of the simple real Lie algebras, except for
$\mf{s}\mf{l}(2,\mb{R})$, have a subalgebra isomorphic to
$\mf{s}\mf{u}(2)$. Hence, this requirement excludes all semisimple algebras
which cannot give rise to a group whose universal cover is not
diffeomorphic to $\mb{R}^n$ by virtue of the Iwasawa decomposition. Using Theorem \ref{thm:Rleq0} the
theorem follows from the above analysis.
\end{proof}
This theorem gives a very precise criterion for which Lie groups admit only negatively curved left-invariant metrics. The existence of an $\mf{s}\mf{u}(2)$ subgroup makes it possible to find a positively curved left-invariant metric on the Lie group. 

\section{Einstein metrics}
A Lie group usually possess many non-isometric left-invariant metrics. However, for a given Lie group, is there a particularly nice or distinguished left-invariant metric? Such metrics are, for example, Einstein metrics for which 
\beq
R_{\mu\nu}=\lambda g_{\mu\nu},
\eeq
where $\lambda$ is a constant. However, what Lie groups admit left-invariant Einstein metrics? Here we will discuss some known  results regarding Einstein metrics on Lie groups. 

Besse \cite{Besse} has devoted a whole book to Riemannian manifolds with Einstein metrics.  
The complete answer for which Lie groups admit such metrics is not known. However, we have the following rough classification \cite{Besse}
\begin{thm}
Let $(M,{\bf g})$ be a homogeneous Einstein manifold, $R_{\mu\nu}=\lambda g_{\mu\nu}$;
\begin{enumerate} 
\item{}if $\lambda>0$, then $M$ is compact with finite fundamental group. 
\item{} if $\lambda=0$, then $M$ is flat.
\item{} if $\lambda<0$, then $M$ is non-compact.
\end{enumerate} 
\end{thm}
A discussion of the positively curved ones is done in \cite{Besse}, and the above theorem implies that the $\lambda=0$ consists of only the flat ones (this was proven in \cite{AK}). Our concern here will be the negatively curved ones, $\lambda<0$. 

From the analysis above we note that the best candidates for Lie groups with Einstein metrics are the solvable groups. However, not all solvable groups allows for an Einstein metric; e.g. Dotti Miatello showed that a solvable unimodular Lie group does not admit a left-invariant Einstein metric of strictly negative curvature \cite{DM}; the only Einstein metric allowed on a solvable unimodular Lie group is the flat one. (Note that a special class of the unimodular solvable groups are the nilpotent ones. This will be of significance a bit later.)

Consider a real Lie algebra, $\mf{s}$, with the following properties:
 \begin{enumerate} 
\item{} The Iwasawa decomposition has the following orthogonal decomposition: 
\[ {\mf s}=\mf{a}\oplus\mf{n},\quad [\mf{s},\mf{s}]=\mf{n},\] 
where $\mf{a}$ is Abelian, and $\mf{n}$ is nilpotent.
\item{} All operators $\mathrm{ad}_{X}$, $X\in\mf{a}$ are symmetric.
\item{} For some $X_0\in\mf{a}$,
  $\left.\mathrm{ad}_{X_0}\right|_{\mf{n}}$ has positive eigenvalues.
\end{enumerate}
Since ${\mf{s}}$ is a solvable algebra, the corresponding group manifolds are so-called
  \emph{solvmanifolds}. 
These geometries are strong candidates for
Einstein spaces with negative curvature \cite{Wolter,Wolter2}. The simplest possible $\mf{s}$ \footnote{The simplest possible $\mf{s}$ is  the Lie algebra defined by $[{ X}_0,{ X}_i]={ X}_i, ~i=1...n-1$, with all other commutators being zero. The corresponding Lie group acts simply transitive on $n$-dimensional real hyperbolic space, $\mb{H}^{n}$.} corresponds to real hyperbolic space, $\mb{H}^{n}$.  
In fact, all known examples of homogeneous Einstein manifolds with negative curvature are of the above type \cite{Heber}. 

Some examples of Einstein solvmanifolds can be found among the symmetric spaces:
\beq
\mb{H}^n&=&SO_0(n,1)/SO(n), \quad \text{(real hyperbolic space)} \nonumber \\ 
\mb{H}^n_{\mb{C}}&=&SU(n,1)/S(U(n)\times U(1)),\quad \text{(complex hyperbolic space)}\nonumber \\
\mb{H}^n_{\mb{H}}&=&Sp(n,1)/Sp(n)\times Sp(1), \quad \text{(quaternionic hyperbolic space)} \nonumber \\
\mb{H}^2_{\sf Cay}&=& F^{-20}_4/\mathrm{Spin}(9), \quad \text{(Cayley hyperbolic plane)} \nonumber 
\eeq
In all of the above examples, the nilpotent algebra $\mf{n}=[\mf{s},\mf{s}]$ is of generalised Heisenberg type. By considering \emph{all} the so-called generalised Heisenberg algebras (which are all two-step nilpotent) we can produce many more examples of such Einstein solvmanifolds \cite{GHG}. For each of the generalised Heisenberg algebras, $\mf{n}$, there is a solvable extension $\mf{s}$ (i.e. $\mf{n}=[\mf{s},\mf{s}]$) whose Lie group admits an Einstein metric. These  solvable extensions of generalised Heisenberg algebras are usually called Damek-Ricci spaces. These Damek-Ricci spaces thus provides us with even more examples of Einstein solvmanifolds. 

More recently, Jorge Lauret developed a new method of producing rank-one (i.e. $\dim(\mf{a})=1$) solvmanifolds \cite{L1,L2,L3}. The method relies on the observation that if $(\mf{n},{\bf g}_{\mf{n}})$ is a so-called \emph{Ricci nilsoliton}, then there is a metric solvable extension $(\mf{s},{\bf g}_{\mf{s}})$ which is Einstein. This result, along with some other useful techniques Lauret developed, reduces the problem of finding rank-one Einstein solvmanifolds to finding Ricci soliton metrics on homogeneous nilmanifolds. Lauret explicitly used his methods to find all rank-one Einstein solvmanifolds up to dimension 6 \cite{L3} while Cynthia Will found all of dimension 7 \cite{Will}. 

\subsection{Inhomogeneous Einstein metrics from Black Holes} The above Einstein solvmanifolds are homogeneous by construction. However, there is a simple way of constructing some inhomogeneous Einstein manifolds using these solvmanifolds as a starting point. The resulting inhomogeneous metrics have the property that they are asymptotically isometric to Einstein solvmanifolds and have an interesting Lorentzian interpretation. 

The paper \cite{Sig} investigated Einstein solvmanifolds and explicitly found some Lorentzian black hole solutions. These black hole solutions are inhomogeneous and incomplete but their inhomogeneous Riemannian counterparts can be made complete and everywhere regular by an appropriate periodical identification. 

By Wick-rotating the black hole solutions in \cite{Sig} -- i.e. $t=i\tau$ -- the solutions can be written
\beq
\d s^2&=& e^{-2pw}F(w)\d \tau^2+\frac{\d w^2}{F(w)}+\sum_i e^{-2q_iw}\left({\mbold\omega}^i\right)^2, \nonumber \\
F(w)&=& 1-Me^{\sigma w},
\eeq  
where $\sigma=p+\sum_iq_i$ and $M>0$.  

The Lorentzian version of this metric describes a higher-dimensional black hole where the horizon geometry are products of solvegeometries and nilgeometries. For example, by considering the solvable extension of a generalised Heisenberg group we get black holes having a generalised Heisenberg group as a horizon. In this way we see that there is a great variety of black hole solutions for negatively curved spaces. 

We do a change of coordinates 
\beq
r^2 = 1-Me^{\sigma w},  \qquad
\tau = \frac{2}{\sigma M^{\frac p\sigma}}\theta,
\label{eq:rtau}\eeq
which transforms the metric into 
\beq
\d s^2=\frac{4}{\sigma^2}\left[\frac{r^2\d\theta^2}{(1-r^2)^{\frac{2p}{\sigma}}}+\frac{\d r^2}{(1-r^2)^{2}}\right]+\sum_i e^{-2q_iw(r)}\left({\mbold\omega}^i\right)^2,
\eeq
where $w(r)$ is  implicitly given via eq.(\ref{eq:rtau}). 

We note that by choosing $\theta$ to be periodic with $0\leq \theta <2\pi$ and $0\leq r <1$, the metric becomes everywhere regular\footnote{Writing $\tau=\beta\theta$, the constant $2\pi\beta$ can be interpreted as the  \emph{inverse temperature of the black hole}; i.e. $2\pi\beta=1/T$. This implies that the temperture is  $T\propto  M^{p/\sigma}$, and hence, $T$ is monotonically increasing in $M$.} . Near the horizon ($r=0$) the metric closes off regularly and hence, the metric describes a complete inhomogeneous Riemannian metric which is Einstein. The mass of the black hole, $M>0$, is arbitrary and parametrises a one-parameter family of inhomogeneous spaces. The spaces are asymptotically  Einstein solvmanifolds as $r\rightarrow 1$.

\section{Ricci Nilsolitons} 
We have already pointed out that nilpotent groups do not allow for a left-invariant Einstein metric; the Ricci tensor, as given in eq.(\ref{eq:Rij}), will for nilpotent groups have both positive and  negative eigenvalues. We thus might wonder whether there are any other ``distinguished'' metrics on nilpotent groups. 

Lauret \cite{L1} noted that some nilpotent groups allow for metrics which obey \beq
R_{\mu\nu}=\lambda g_{\mu\nu}+D_{\mu\nu},
\label{eq:nilsoliton}\eeq
where $D^{\mu}_{~\nu}$ as a linear map, ${\sf D}:\mf{n}\mapsto\mf{n}$, is a \emph{derivation of $\mf{n}$}; i.e. 
\[ {\sf D}\left([{ X},{ Y}]\right)=[{\sf D}({X}),{ Y}]+[{ X},{\sf D}({ Y})]. \]
These metrics have a nice interpretation in terms of special solutions of the \emph{Ricci flow} \cite{Hamilton}. For a curve ${\bf g}(t)$ of Riemannian metrics on a manifold $M$, the Ricci flow is defined by the equation 
\beq
\frac{\partial g_{\mu\nu}}{\partial t}=-2R_{\mu\nu}.
\label{eq:Ricciflow}\eeq
If a solution to the Ricci flow (\ref{eq:Ricciflow}) moves by a diffeomorphism and is also scaled by a factor at the same time, we call the solution a \emph{homothetic Ricci soliton}. In other words, if $\phi_t$ is a one-parameter family of diffeomorphims generated by some vector field, then if there is a solution of the form 
\[ {\bf g}(t)=c(t)\phi_t^*{\bf g}, \]
then ${\bf g}$ is  a homothetic Ricci soliton. 

Ricci nilsolitons are nilmanifolds with left-invariant metrics allowing for such solutions. 
Moreover, by inspection, all Einstein metrics will be so too. Hence, this is a way of 
generalising the Einstein requirement to nilpotent groups. These Ricci nilsolitons are unique 
up to isometry and scaling and can therefore be taken to be distinguished left-invariant metrics 
on nilmanifolds. 

Lauret also showed the following property
\begin{thm}[Lauret \cite{L1}]
A homogeneous nilmanifold $(\mf{n},{\bf g}_{\mf{n}})$ is a Ricci nilsoliton if and only if $(\mf{n},{\bf g}_{\mf{n}})$ admits a metric solvable extension $(\mf{s},{\bf g}_{\mf{s}})$ ($\mf{s}=\mf{a}\oplus\mf{n}$) with $\mf{a}$ Abelian whose corresponding solvmanifold is Einstein. 
\end{thm}
Hence, this result intertwines the problem of finding Einstein solvmanifolds and finding Ricci nilsolitons. 

Furthermore, Lauret also noted that the Ricci nilsolitons are fixed points of the functional
\beq
{\bf g}\mapsto S[g_{\mu\nu}]=\int_M R_{\mu\nu}R^{\mu\nu}\sqrt{g}\d^nx,
\eeq
restricted to a certain subset (basically the unit circle) of the space of all nilpotent brackets\footnote{Details of this variational procedure can be found in his papers \cite{L1,L2,L3,L4}.}. 

Examples of such Ricci nilsolitons are easy to find. For example, all metric groups of generalised Heisenberg type are Ricci nilsolitons. Moreover, all nilpotent groups of dimension 6 or lower, admit a Ricci nilsoliton metric (see \cite{Will} where all the 6-dimensional ones are given). 

\subsection{Higher-curvature gravitational nil-instantons} 
Many fundamental theories of physics are defined via an action principle. For example, Einstein gravity can be defined through the Einstein-Hilbert action:
\beq
{\bf g}\mapsto S_{EH}[g_{\mu\nu}]=\int_M (R-2\Lambda)\sqrt{|g|}\d^nx.
\eeq
By requiring that the metric should be a fixed point of this map, implies the Einstein equation to be satisfied:
\beq
R_{\mu\nu}-\frac 12 Rg_{\mu\nu}+\Lambda g_{\mu\nu}=0.
\label{eq:Einstein}\eeq
Compact\footnote{A nilmanifold can be compactified if and only if there exists a frame such that $[e_i,e_j]=C^k_{ij}e_k$, where $C^k_{ij}$ are all rational constants \cite{Eberlein}.} Riemannian manifolds (with a totally geodesic boundary) which are solutions to this equation are usually called \emph{gravitational instantons}. 

It is easy to see that the Einstein equation (\ref{eq:Einstein}) is equivalent to saying that the metric is Einstein; i.e. $R_{\mu\nu}=\lambda g_{\mu\nu}$. We noted earlier that nilmanifolds do not admit an Einstein metric so there will be no non-trivial nilmanifolds being solutions to this equation. However, we will in the following  rephrase the problem and ask: Are there actions for which the Ricci nilsolitons are solutions? 

The result of Lauret gives us some hints of what such  actions look like; they seem to  contain quadratic terms like $R_{\mu\nu}R^{\mu\nu}$. Let us therefore consider a class of higher-curvature gravity theories for which the action takes the form 
\beq
{\bf g}\mapsto S_{(\alpha,\beta)}[g_{\mu\nu}]=\int_M \left(R+\alpha R^2+\beta R_{\mu\nu}R^{\mu\nu}-2\Lambda\right)\sqrt{g}\d^nx.
\eeq
The variation of the above action implies that the metric have to obey the generalised Einstein equation:
\beq
\Phi_{\mu\nu}+\Lambda g_{\mu\nu}=0,
\label{eq:GenEinstein}\eeq
where 
\beq
\Phi_{\mu\nu}&=&R_{\mu\nu}-\frac 12 Rg_{\mu\nu}+2\alpha R\left(R_{\mu\nu}-\frac 14 Rg_{\mu\nu}\right) \nonumber \\
&& +(2\alpha+\beta)(g_{\mu\nu}\Box-\nabla_{\mu}\nabla_{\nu})R+\beta\Box\left(R_{\mu\nu}-\frac 12 Rg_{\mu\nu}\right) \nonumber \\
&& +2\beta\left(R_{\mu\sigma\nu\rho}-\frac 14g_{\mu\nu}R_{\sigma\rho}\right)R^{\sigma\rho}, 
\eeq
and $\Box\equiv \nabla^{\mu}\nabla_{\mu}$. 

First note that all Einstein manifolds, $R_{\mu\nu}=\lambda g_{\mu\nu}$, are solutions to eq.(\ref{eq:GenEinstein}); hence, in some sense, this equation generalises the Einstein equation. 

Regarding the nilsolitons, we  also note that they are solutions for particular values of the parameter values $(\alpha,\beta,\Lambda)$. For example, consider the four-dimensional nilsoliton, ${\sf Nil}^4$:
\beq
\d s^2=\d w^2+\left(\d x-a y\d w\right)^2+\left(\d y-a z\d w\right)^2+\d z^2,
\eeq
where $a$ is a constant. 
This is a solution to eq.(\ref{eq:GenEinstein}) if and only if 
\beq
{\sf Nil}^4: \quad 2\alpha+3\beta=\frac{1}{a^2}, \quad \Lambda=-\frac{a^2}{4}.
\eeq
Interestingly, the action for this solution is zero: $S_{(\alpha,\beta)}[{\sf Nil}^4]=0$. 

Let us consider some other examples. We consider the generalised Heisenberg groups with the appropriate nilsoliton metric, $\mc{H}_{m,n}$. In Heber's notation \cite{Heber}, the metric solvable extension of this group admits an Einstein metric of eigenvalue type $(1<2;n,m)$. This means that $\dim \mc{H}_{m,n}=m+n$. 
Note that  equation (\ref{eq:GenEinstein}) is invariant under a  rescaling of the metric $\d s^2\mapsto \ell^2\d s^2$, along with a simultaneous rescaling of the parameters: $(\alpha,\beta,\Lambda)\mapsto (\ell^2\alpha,\ell^2\beta,\ell^{-2}\Lambda)$. After an appropriate rescaling, the spaces  $\mc{H}_{m,n}$ are solutions to eq.(\ref{eq:GenEinstein}) if 
\beq
\mc{H}_{m,n}:\quad \alpha+\frac{n+4m}{nm}\beta=\ell^2, \quad \Lambda=-\frac{1}{8\ell^2}. 
\eeq
Also for these solutions, the action is zero: $S_{(\alpha,\beta)}[\mathcal{H}_{m,n}]=0$. 

Thus it seems that the nilsolitons are solutions of certain higher-curvature theories of gravity.  Admittedly, the approach we have followed here is in some sense the opposite of what is common. We have a set of spaces with particularly nice metrics, then we are trying to find the appropriate action which has these as solutions. 

We should also point out what happens for other types of Lie algebras. There are also metrics on solvable Lie groups being solutions of eq.(\ref{eq:GenEinstein}). Also in these cases eq.(\ref{eq:GenEinstein}) seems to pick out a particularly nice metric. Regarding the semisimple groups, assuming $\beta>0$, implies that the metric must be Einstein. 

\section{Outlook}
In low-dimensional topology and geometry the homogeneous spaces seem to play an important role. 
According to a conjecture by Thurston \cite{thurston}, any compact 3-manifold can be decomposed 
into primes where each prime has to be one of 8 geometries. These 8 geometries are all homogeneous 
and are called ``model geometries''\cite{thur:97}. Model geometries are special classes of homogeneous
spaces and may be more interesting in a topological context. One of the biggest problems of proving Thurston's 
conjecture for 3-manifolds is the enormous variety and richness of compact hyperbolic 3-manifolds \cite{BP}. 
In spite of the fact that they seem to be the most numerous, these negatively curved spaces seem 
to be the least understood of the 8 model geometries. 

Another set of "distinguished metrics" are \emph{bi-invariant metrics}. For semi-simple groups, the bi-invariant 
metrics are related to Einstein metrics \cite{Milnor:76}, and allowing for arbitrary signature, the 
relation is even more striking. For an investigation of bi-invariant metrics (not necessary Riemannian) 
on Lie groups, see, for example, \cite{Muller}.   

Another thing it would be interesting to investigate is the relation between the symmetries of the negatively curved spaces and the symmetries of their conformal boundaries. For the simplest cases, namely the real hyperbolic spaces, we have the amazing correlation: $\mathrm{Isom}(\mb{H}^n)=\mathrm{Conf}(\partial \mb{H}^n)$ (see e.g. \cite{BP}). For $\mb{H}^n$ the isometries in the interior are uniquely determined by their action the conformal boundary. This interplay between the structure in the interior and on the conformal boundary ultimately gives rives rise to the celebrated AdS/CFT correspondence in theoretical physics. One might contemplate whether a similar correspondence holds for other negatively curved spaces. 

This work has entirely been devoted to Riemannian manifolds; however, Lorentz\-ian manifolds are 
equally interesting, particularly for theoretical physicists. Classifying the Lorentzian manifolds 
prove to be harder than the Riemannian case because some of the uniqueness results fail for 
Lorentzian manifolds. For example, homogeneous Riemannian manifolds are uniquely characterised by 
their curvature invariants \cite{PTV}; however, for Lorentzian manifolds even after requiring that 
the metric to be homogeneous and Einstein, the curvature invariants do not uniquely determine the 
metric \cite{solwaves}. Notwithstanding, due to the role Lorentzian spaces play in fundamental theories 
of our physical universe, we believe that understanding the Lorentzian case would be extremely valuable. 

Here we have only given a flavour of the enormous variety of geometrical structure these 
negatively curved homogeneous spaces have to offer. More study is clearly required to fully 
uncover and understand their geometric properties. Maybe in the future their apparently almost 
unlimited potential will be fully appreciated.

\section*{\textit{Note added:}}
Since these notes were written (June 2004), there have been further developments in the field. In particular, Lauret \cite{L07} has showed that all Einstein solvmanifolds are indeed standard, i.e., of the form described in section 5 (see also, \cite{L08}). Furthermore, the existence of the higher-curvature nil-instantons led me to realise that higher-curvature theories of gravity have a peculiar set of solutions which, from a cosmological standpoint, inflate anisotropically \cite{BH1} (in fact, these exact anisotropically inflating solutions are solvmanifolds). This may have interesting consequences for early-universe cosmology and the cosmic microwave background \cite{BH2,BH3}.  Moreover, the inhomogeneous Einstein metrics obtained from the black hole solutions were also discussed in \cite{NilBH}.

\section*{Acknowledgments}
I would like to thank the organisers for letting me talk in the CMS Summer 2004 meeting in Halifax, Nova Scotia. Thanks also to D. Andriot for pointing out an error in the first version of the draft.
This work was funded by the Killam Trust and AARMS.

\end{document}